\documentclass[aps,prl,reprint,superscriptaddress]{revtex4-1} 
\usepackage{graphicx}
\usepackage{xcolor}
\usepackage{textcomp}
\begin{document}

\title{The detection of Kondo effect in the resistivity of graphene: artifacts and strategies }

\author{Johannes Jobst}
\author{Ferdinand Kisslinger}
\author{Heiko B.\ Weber}
\email{heiko.weber@physik.uni-erlangen.de}
\homepage{http://www.lap.physik.uni-erlangen.de/}

\affiliation{Lehrstuhl f\"ur Angewandte Physik, Universit\"at
  Erlangen-N\"urnberg, 91056 Erlangen, Germany}

\date{\today}

\begin{abstract}

We discuss the difficulties to discover Kondo effect in the resistivity of graphene. Similarly to the Kondo effect, electron-electron interaction effects and weak localization appear as logarithmic corrections to the resistance. In order to disentangle these contributions, a refined analysis of the magnetoconductance and the magnetoresistance is introduced. We present numerical simulations which display the discrimination of both effects. Further, we present experimental data of  magnetotransport. When magnetic molecules are added to graphene, a logarithmic correction to the conductance occurs, which apparently suggests Kondo physics. Our thorough evaluation scheme, however, reveals that this interpretation is not conclusive: the data can equally be explained by electron-electron interaction corrections in an inhomogeneous sample. Our evaluation scheme paves the way for a more refined search for the Kondo effect in graphene.
\end{abstract}

\maketitle


The Kondo effect is one of the most intriguing effects in condensed matter physics \cite{kondo}. It is a consequence of the many-body interaction of magnetic degrees of freedom with the conduction electrons in a metal. We are particularly interested in Kondo effect in graphene, which is expected to be different from Kondo effect in conventional metals due to the specific structure of the quantum mechanical electronic wave function \cite{vojta}. So far, there is no convincing experimental evidence for Kondo effect in graphene. We consider an earlier apparent finding of Kondo effect as a misinterpretation \cite{fuhrer-kondo, jobst-correspondence, fuhrer-correspondence}, the origin of which we elaborate in this manuscript. 
Graphene, in contrast to conventional (buried) two-dimensional electron gases, provides a unique opportunity to add  magnetic degrees of freedom and couple them to the electronic system. They may be added by: (i) structural defects in the graphene layer which are identified as magnetic impurities \cite{geim-magnetic-defects, fabian}, (ii) states with unpaired electrons are present at monolayer/bilayer interfaces, and in dangling-bond states of the substrate \cite{pankratov}, (iii) magnetic ions or molecules may be added to the surface of graphene \cite{manoharan, wehling, fabian}. 
This does not mean, however, that Kondo physics is conveniently accessible by experiments: In the low-density electronic system graphene, Kondo temperatures may be extremely small \cite{vojta} and thus, the effect may be shifted to a temperature regime which is experimentally not accessible. However, with sufficiently strong exchange coupling, and remote from the Dirac point, convenient Kondo temperatures are conceivable. 


A further experimental difficulty is the unambiguous identification of Kondo effect. The typical trace would be a logarithmic increase of the resistivity $\rho(T)$ towards low temperatures $T$, the amplitude of which scales with the density of magnetic impurities. This increase saturates at even lower temperatures when the magnetic impurities are fully screened by the conduction electrons. 
However, in graphene the two-dimensionality causes the situation that three logarithmic-in-T quantum corrections to the resistivity occur: weak localization (WL), electron-electron interaction correction (EEI), and Kondo physics. 
This leads to the case that very similar $\rho(T)$ signals have been interpreted in the framework of EEI \cite{camassel-EEI, lara-avila} and Kondo physics \cite{fuhrer-kondo}, respectively. Note that WL can reliably be suppressed in finite magnetic fields.

Epitaxial graphene grown on the (0001) face of semi-insulating silicon carbide (SiC) \cite{emtsev-natmat, jobst-QHE} is particularly well suited for the search for Kondo effect. It provides a large graphene sheet with a rather high electron density $n\approx 10^{13}$\,cm$^{-2}$. It further has displayed a parameter-free agreement with predictions for EEI corrections \cite{jobst-EEI}, which is due to the excellent lateral homogeneity of the material. 

In this letter we present a method to disentangle Kondo effect and EEI. We will discuss experimental data in order to display the uncertainty in the identification of Kondo effect. We then provide a scheme for refined data analysis, which discriminates the two logarithmic contributions excellently, at least for perfectly homogeneous samples. Finally, we will demonstrate how inhomogeneous sample parameters apparently generate Kondo signals and thus cause misinterpretations.

\begin{figure}[htb]
  \centering
	\includegraphics[width=86mm]{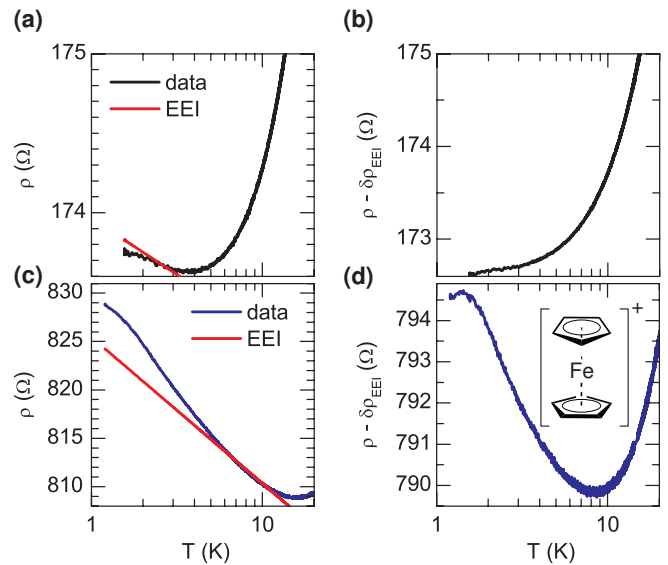}
	\caption{(a) A logarithmic temperature dependence of the resistivity is observed in as prepared graphene. (b) After subtracting the expected EEI correction (red line) the logarithmic dependence vanishes, leaving the phonon-dominated resistivity. (c) Deposition of ferrocenium ions [see inset in (d)] on the graphene surface enhances the amplitude of the logarithmic corrections notably. (d) In addition, a Kondo-like logarithmic feature remains after the expected EEI correction is subtracted.  \label{fig:ferrocenium}}
\end{figure}

We have prepared large-area Hall bars (channel width $b=50$\,\textmu m, channel length $l=330$\,\textmu m) from epitaxial graphene. Figure \ref{fig:ferrocenium}(a) displays the longitudinal resistivity $\rho_{xx}(T)$ in a magnetic field of $B=0.5$\,T, which suppresses the WL contribution. Data are obtained from low-frequency lock-in, four-terminal measurements. We find a Drude resistivity $\rho_0 = 1/(en\mu) \approx 173\,\Omega$ with $n=1.2\cdot10^{13}$\,cm$^{-2}$ and a Drude mobility $\mu=3000$\,cm$^2$/Vs. Moreover, we observe a weak logarithmic increase of $\rho_{xx}(T)$ towards low temperatures. 
The conductance tensor of a two-dimensional metal reads:
\begin{equation}
	\hat{\sigma} = \frac{en\mu}{1 + \mu^2 B^2}\left(
\begin{array}{cc}
	1 & -\mu B\\
	\mu B & 1
\end{array}
\right) +
\left(
\begin{array}{cc}
	\delta\sigma_{\text{EEI}} & 0\\
	0 & \delta\sigma_{\text{EEI}}
\end{array}
\right)
\label{eq:sigma-tensor}
\end{equation}
with the logarithmic-in-$T$ EEI correction $\delta\sigma_{\text{EEI}}$ \cite{altshuler} which acts only on the diagonal terms. By inversion of $\hat\sigma$, one obtains the resistivity tensor. Its diagonal term 
%
\begin{equation}
	\rho_{xx} = \rho_0 + [\mu^2 B^2 - 1]
        \frac{e^2 \rho^2_0}{2\pi^2 \hbar} \left[A \cdot
          \ln\left(\frac{k_{\text{B}} T \tau_{\text{tr}}}{\hbar}\right) \right]
        \text{,}
	\label{eq:EEI}
\end{equation}
%
describes both, the logarithmic temperature dependence, and the parabolic magnetic field dependence of  $\rho_{xx}$.
For graphene on SiC $A=0.86$ is theoretically expected and experimentally confirmed at low temperatures \cite{jobst-EEI}. The only free parameter is $\rho_0$. It turns out that this description is reasonable but slightly overestimates the logarithmic increase in Fig.\ \ref{fig:ferrocenium}(a). The difference between EEI predictions and raw data is displayed in Fig.\ \ref{fig:ferrocenium}(b). It is presumably a consequence of the strong $\rho(T)$ in this material which is typically assigned to phonon scattering \cite{hibino-SPP, speck-QFMLG}. 

Next, magnetic scatterers are added to the graphene surface by dropcasting. We opted for ferrocenium molecules, where the central Fe$^{2+}$-ion provides an unpaired electron state which extends over the whole molecule. 
Subsequently, the low-temperature measurements are repeated. Although we are convinced that this treatment did not damage the graphene layer or the SiC substrate chemically, a  very strong change in $\rho_{xx}(T)$ is observed (see Fig.\ \ref{fig:ferrocenium}(c)). First, the overall resistivity of the very same sample has increased roughly by a factor of five. Second, the logarithmic increase has become more pronounced. In absolute resistivity values it now counts 20\,$\Omega$ compared to 0.1\,$\Omega$ in Fig.\ \ref{fig:ferrocenium}(a). When subtracting the expected EEI correction (eq.\ \ref{eq:EEI}), the remaining signal looks perfectly like the targeted Kondo feature. Moreover, its amplitude increases when applying more and more molecules (not shown). 
One is thus tempted to assign this experimental result as the appearance of Kondo physics once magnetic scatterers are added. It will be shown, however, that this conclusion should be taken \emph{cum grano salis}. A more careful analysis will show that sample inhomogeneities, in conjunction with EEI, cause very similar phenomena. 

To understand this we propose a Gedankenexperiment. When adding a magnetic impurity to graphene it may possibly give a logarithmic correction to $\rho$ in the framework of Kondo physics (dynamic impurity scattering). However, it also adds a new static scattering center and thus enhances $\rho_0$. Via equation \ref{eq:EEI}, also the logarithmic EEI correction is increased. As a consequence one added magnetic impurity contributes to the logarithmic correction twofold: potentially as Kondo scatterer, but unavoidably via EEI.

We now turn back to the Kondo-like difference in Fig.\ \ref{fig:ferrocenium}(d). This signal is potentially due to Kondo physics, or due to an inaccurate estimate of $\rho_0$. The low-temperature saturation seems to play in favor of the Kondo scenario, but could also be an experimental problem of insufficient thermalization at low $T$. Note that this curve is very similar to the data presented as Kondo effect in Ref.\ \cite{fuhrer-kondo}, where however, EEI was completely disregarded. 

We propose a refined data analysis. For the discrimination of Kondo and EEI corrections, it is useful to recall their origin. Kondo effect is a renormalization of the scattering time $\tau$ and therefore acts on the resistivity. EEI, however, is a correction to the conductivity. This difference can be used to disentangle both effects. 
The analysis of the curvature of the parabolic magnetoresistance $\rho_{xx}(B,T)$ is used to quantify EEI corrections, but is insensitive to Kondo contributions. In analogy, the shape of the magnetoconductance $\sigma_{xx}(B,T)$ is insensitive to EEI but gives access to the Kondo correction. 
Roughly, its curvature indicates the temperature dependence of Kondo scattering. A more thorough treatment extracts a temperature-dependent $\mu(T)$ from fitting Eq.\ \ref{eq:sigma-tensor} to magnetoconductivity data recorded at various temperatures. 
\begin{figure*}[htb]
  \centering
	\includegraphics[width=178mm]{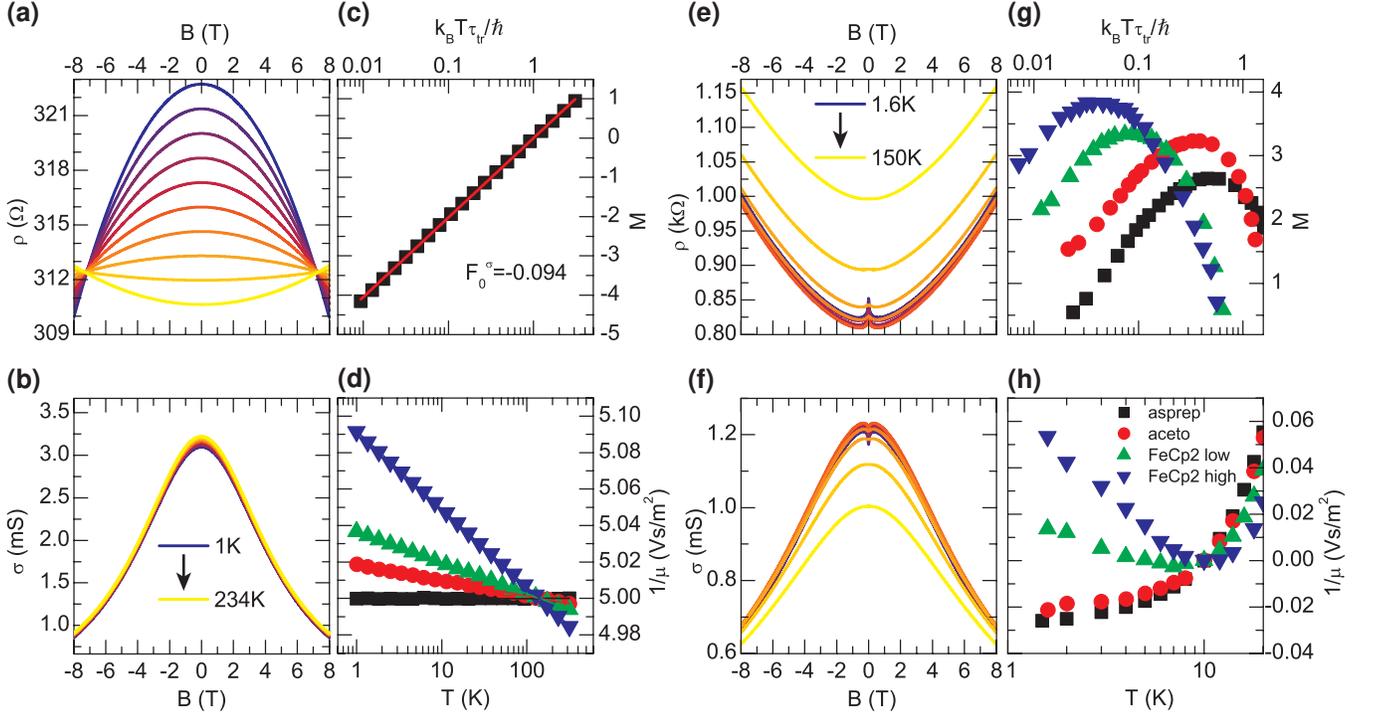}
	\caption{(a--d) The evaluation scheme is tested on an artificially created data set with $n=10^{13}$\,cm$^{-2}$, $\mu = 2000$\,cm$^2$/Vs and $\rho_{\text{K}} = 5\,\Omega$ in a temperature range from 1\,K to 234\,K. The evaluation of the curvature of the magnetoresistivity in (a) yields the EEI contribution only (c). Analyzing the shape of the magnetoconductivity in (b) yields the magnitude of the Kondo correction (d). In (d) the evaluation is shown for data sets without Kondo correction (black squares), and for $\rho_{\text{K}} = 1\,\Omega$ (red circles), $\rho_{\text{K}} = 2\,\Omega$ (green upwards triangles) and $\rho_{\text{K}} = 5\,\Omega$ (blue downwards triangles). The evaluation is in perfect agreement with the set $\rho_{\text{K}}$ values. 
	(e--h) Experimental data for a sample before and after ferrocenium deposition. (e,f) show the magnetoresistivity and conductivity for the high ferrocenium concentration, respectively. (g) The evaluation of the $\rho_{xx}(B,T)$ data reliably separates the EEI correction for all ferrocenium concentrations. (h) The Kondo contribution that is calculated from the shape of $\sigma_{xx}(B,T)$ is $\rho_{\text{K}} = -2.1\,\Omega$, $\rho_{\text{K}} = -1.6\,\Omega$, $\rho_{\text{K}} = 3.7\,\Omega$, $\rho_{\text{K}} = 13.3\,\Omega$ for the as-prepared sample, after dropcasting of pure acetonitrile, with a ferrocenium concentration of $[\text{FeCp}_2^+] \approx 2\cdot10^{10}$\,cm$^{-2}$ and $[\text{FeCp}_2^+] \approx 8\cdot10^{11}$\,cm$^{-2}$, respectively.
	\label{fig:magneto}}
\end{figure*}       

This evaluation scheme is tested with an artificially created data set where the Kondo contribution can be parametrically varied. Therefore, we set $n=10^{13}$\,cm$^{-2}$ and $\mu=2000$\,cm$^2$/Vs. This fixes the Drude conductivity and the EEI correction. In a next step, a Kondo correction is added to $\rho_{xx}$ as
\begin{equation}
	\delta\rho_{\text{K}}(T) = \frac{0.47}{2}\rho_{\text{K}}\ln(T)
\label{eq:Kondo}
\end{equation}
with $\rho_{\text{K}}$ varied from 0 to 10\,$\Omega$. These values provide a resistivity correction of similar magnitude as the EEI correction $\delta\rho_{\text{EEI}}$. Figure \ref{fig:magneto} shows the magnetoresistivity (a) and conductivity (b) calculated from this data set for various temperatures and with $\rho_{\text{K}}=5\,\Omega$. 
The evaluation of the magnetoresistivity following the route in \cite{jobst-EEI} is shown in Fig.\ \ref{fig:magneto}(c). The theoretically expected $A=0.86$ is found irrespective of the magnitude of the included Kondo correction. This displays that this evaluation scheme of the magnetoresistivity projects out the EEI correction only. 
In analogy, the magnetoconductance data are evaluated and deliver a logarithmic dependence of $1/\mu$ on $T$. The magnitude exactly reflects the input values for all values of $\rho_{\text{K}}$. This is not surprising, but validates that the data evaluation procedure discriminates Kondo and EEI contributions reliably, at least under ideal conditions. 

This procedure is now applied to the experimental magnetotransport data obtained with the ferrocenium molecule. The magnetoresistance data look quite different because they include a WL peak around $B=0$, a classical ($T$-independent) magnetoresistivity contribution and a strong electron-phonon scattering. Note that the latter was absent in Ref.\ \cite{jobst-EEI} because there quasi-freestanding monolayer graphene was chosen \cite{speck-QFMLG}. Nevertheless, the evaluation of the $\rho_{xx}(B,T)$ delivers a logarithmic increase in Fig.\ \ref{fig:magneto}(g) which has the expected slope, indicating quantitative agreement with the EEI correction. 
For the analysis of Kondo physics, we now evaluate magnetoconductance data (Fig.\ \ref{fig:magneto}(f)). As an overall impression, the data look very similar to the idealized, generated data set in Fig.\ \ref{fig:magneto}(b). Kondo effect would occur as a logarithmic $1/\mu(T)$ dependence. Indeed, a very small logarithmic signal can be extracted (Fig.\ \ref{fig:magneto}(h)). This signal increases with the ferrocenium concentration. Again, this result based on the refined evaluation scheme, apparently indicates Kondo physics. 

Due to the dropcasting process of applying the ferrocenium ion, homogeneous conditions on the whole sample can not be guaranteed. Consequently, we now analyze the impact of inhomogeneous parameters on the evaluation scheme, both in experiments being described in SI, and in well-controlled simulations. To keep the simulation transparent and to stress the essence of the impact of inhomogeneity, we assume the sample being split into two areas with different parameters. 

\begin{figure}[htb]
  \centering
	\includegraphics[width=86mm]{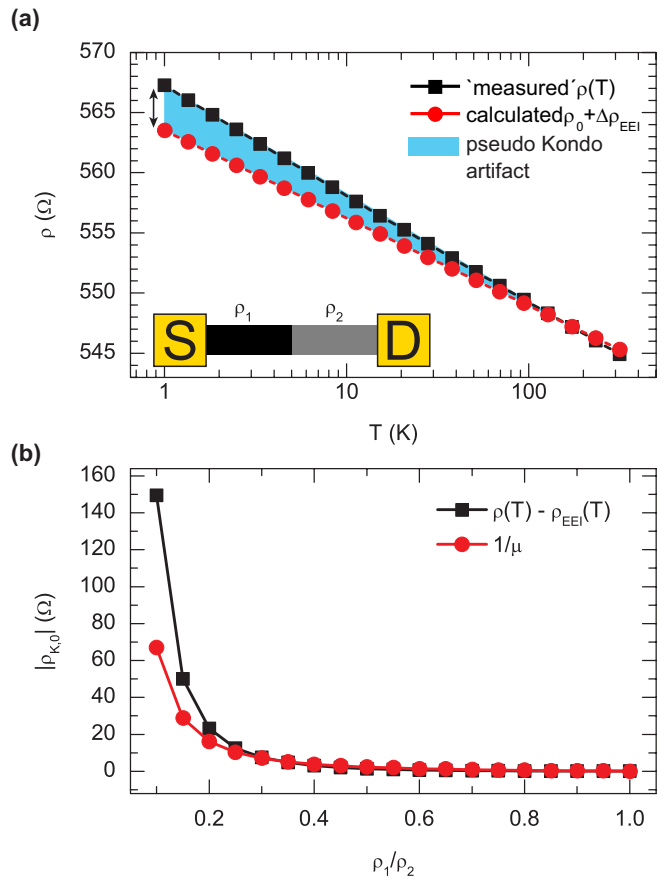}
	\caption{If two areas with different $\rho_0$ are connected in series an apparent Kondo feature arises due to the EEI correction. (a) The apparent EEI (red circles) that is calculated from the \lq measured\rq\ $\rho_0$ is always smaller than the true EEI contribution (black squares). Thus, a pseudo Kondo artifact arises (shaded area). Here an example with $\rho_1/\rho_2=0.4$ and $\rho_{\text{K,}0}$ is shown. (b) The magnitude of the pseudo Kondo artifact grows with decreasing resistivity ratio for both, the evaluation of the longitudinal resistivity (black squares) and the magnetoconductivity (red circles). \label{fig:simulation}}
\end{figure}

We first discuss the case of two areas with different $\rho_1$ and $\rho_2$ in series, as displayed in the inset in Fig.\ \ref{fig:simulation}(a). A Hall measurement (assuming homogeneity) would lead to wrong value of $\rho_{0,m} = (\rho_1 + \rho_2)/2$. If we calculate the EEI expectation from this apparent $\rho_{0,m}$ we would find a contribution as displayed in Fig.\ \ref{fig:simulation}(a) as red line. A \emph{measurement} of $\rho_{xx}(T)$ would add EEI corrections from both regions and would deliver the black curve, which is also logarithmic, but larger in amplitude. One would recognize an unexplained logarithmic contribution (shaded area in Fig.\ \ref{fig:simulation}(a)) which may be attributed to Kondo physics. In this simulation, where Kondo physics is absent, it is only a consequence of EEI plus inhomogeneity. Note that any choice of $\rho_1 \neq \rho_2$ results in an underestimation of EEI and thus creates a \emph{pseudo Kondo artifact}. A closer look reveals that it is unimportant whether the inhomogeneity is caused by different charge densities or charge carrier mobilities. If we assume Kondo physics and a correction following Eq.\ \ref{eq:Kondo}, it just adds a further logarithmic contribution.
 
One may assume that the refined evaluation scheme presented above resolves the EEI and Kondo contributions much better than the simple analysis of $\rho_{xx}$. In analogy to the evaluation procedure that lead to Fig.\ \ref{fig:magneto}(d), we analyze the simulated magnetoconductance data of an inhomogeneous sample. Even in the absence of any Kondo term, we always find a logarithmic $T$-dependence in $1/\mu$ as long as $\rho_1 \neq \rho_2$, which may easily be misinterpreted as Kondo physics. Hence, the apparent Kondo effect found in Fig.\ \ref{fig:magneto}(h) could simply be caused by EEI and inhomogeneity. It can be quantified by parameterizing the inhomogeneity by the ratio $\rho_1/\rho_2$ (Fig. \ref{fig:simulation}(b)). 
It should also be noted that upon adding impurities to graphene, the weak localization peak becomes broader. Hence, an analysis that compares the logarithmic contributions before and after adding defects at the very same magnetic field \cite{fuhrer-kondo} further collects logarithmic contributions from WL. 

In the light of these findings, one may discuss the best strategy to disentangle Kondo physics from WL and EEI. One direction would be to focus on very small sample areas, which set a low-momentum cutoff (saturation) to WL and EEI at low temperatures. Hence, the Kondo effect may be singled out. The graphene sample is in this case, however, very sensitive to edge disorder, localization  and mesoscopic fluctuations, which again create an unclear situation. We favor the opposite strategy that is the use of large-area samples, for which the logarithmic contributions WL and EEI are well controlled and can be separated \textit{via} the presented evaluation scheme of the magnetoconductance. The above-mentioned scenarios, however, elucidate that for this strategy homogeneity is crucial to avoid artifacts that could be misinterpreted as Kondo effect. After all, for robust conclusions in this delicate situation additional probes are mandatory. Obviously, local spectroscopic information obtained with scanning tunneling spectroscopy, together with a careful analysis of the nonlocal logarithmic resistivity would be the best combination for an unambiguous identification of the Kondo effect. For the experiments presented in this manuscript, this could not yet be done, hence an unambiguous assignment to either Kondo physics or to EEi plus inhomogeneity is not yet possible.

To conclude, when adding magnetic scatterers to graphene, an additional logarithmic-in-T contribution to the resistance occurs. Before assigning this to Kondo physics, significant care should be taken. Not only the dynamical Kondo scattering, but also the static scatterer adds \textit{via} EEI logarithmic corrections to the longitudinal resistance. We propose an evaluation scheme that analyzes the magnetic field dependence of the conductance tensor. Under ideal conditions this reliably separates the Kondo and EEI contributions. This separation is, however, susceptible to macroscopic sample inhomogeneities, as demonstrated in experiment and simulation. Our evaluation scheme helps to identify artifacts and paves the way for a more refined search for the Kondo effect in graphene.

The work was carried out in the framework of the Sonderforschungsbereich 953 \textit{Synthetic carbon allotropes}. We acknowledge discussions with Igor Gornyi, Alexander Mirlin, Karsten Meyer and Andreas G\"orling.

\bibliography{Kondo_simulations}

\end{document}


\title{The detection of Kondo effect in the resistivity of graphene: artifacts and strategies -- Supplementary Information}

\author{Johannes Jobst}
\author{Ferdinand Kisslinger}
\author{Heiko B.\ Weber}

\affiliation{Lehrstuhl f\"ur Angewandte Physik, Universit\"at
  Erlangen-N\"urnberg, 91056 Erlangen, Germany}

\date{\today}

%

\maketitle
\vspace{10cm}

In the main manuscript it was discussed that adding an impurity to a graphene sheet adds a logarithmic correction to the resistivity \textit{via} the electron-electron interaction (EEI) correction to the conductivity. In addition, when the sample is inhomogeneous, but the standard evaluation scheme assuming homogeneity is applied, an extra  logarithmic--in--$T$ contribution in the resistivity occurs, which may wrongly be assigned to Kondo effect. In order to demonstrate this, we carried out simulations (in the main manuscript) and experiments (here). We opted for the simplest conceivable model: two areas of monolayer graphene in series, with two different $\rho_0$ (see sketch in Fig. S1(a)). The left (black) part has the resistance $\rho_{0,l}$ and the right (grey) part was irradiated with Ar$^+$--Ions until the resistance doubled $\rho_{0,r} \approx 2 \rho_{0,l}$ at room temperature (at lower $T$, this ratio even becomes larger). Technically, this was achieved by protecting the left part with a PMMA mask during ion exposure. This experiment reproduces the simulation presented in the main manuscript, with $\rho_{0,l}$ and $\rho_{0,r}$ corresponding to $\rho_{1}$ and $\rho_{2}$ in the simulation, respectively. \\

We now report resistivity measurements of several sample areas, first of the undamaged area (Fig. S1 (a)), then of the intentionally damaged area (b), and subsequently of the overall sample (c). The data were collected at a magnetic field of 1$\,$T where weak localization (WL) does not play a significant role. Quantitatively, the WL contribution varies only about 1$\,\mathrm{\Omega}$ in the considered temperature interval (up to $30\,$K). 
The resistivity of the undamaged area is composed of a high-temperature contribution of electron-phonon scattering, and a logarithmic increase towards lower temperatures. Next, the EEI contribution (see Eqn. 2 in the main manuscript) is fitted to the data with  $\rho_0$ as the only free parameter. The EEI theory accurately and completely describes the low temperature behavior. When reproducing the same measurements in the damaged area (see Fig. S1 (b)), the overall resistivity is significantly higher. When doing the same procedure, EEI theory matches roughly, but now a small discrepancy is observed, which we will discuss later. 
The comparison with the simulation presented in Fig. 3 of the main manuscript is the data shown in Fig. S1c, where the resistivity measurement is carried out on the entire sample, disregarding the inhomogeneity. The data thus obtained have first of all a wrong value of the resistivity, which results from the addition of resistors in series ($\rho_0 = (\rho_{0,l} + \rho{0,r})/2$). A more careful analysis reveals that the areas split rather like 52\,\% undamaged and 0.48\,\% damaged area.

More importantly, there is an extra logarithmic contribution showing up, which corresponds to the shaded area in Fig. 3(a) of the main manuscript. In this well-controlled experiment, it is obviously a consequence of EEI plus sample inhomogeneity. However, when the inhomogeneity is not considered, the discrepancy (pseudo-Kondo artifact) would likely be assigned to Kondo physics \cite{fuhrer-kondo}. Figure S1(d) displays the resistivity correction  $\rho (T)-\rho_{EEI}(T)$ derived by subtracting the EEI expectations of the data in Fig. S1(c). As we have measured both areas independently, we can resolve two contributions: the dark shaded area stems from EEI and inhomogeneity. This is the same effect that is simulated in the main manuscript and is what we termed pseudo-Kondo-artifact. 

In addition, a second contribution arises in the experiment (light shaded area), which is absent in the simulations. It corresponds to the deviation that became visible in Fig. S1(b). There are two possible explanations. First, it could be Kondo effect generated by local structural defects. Second (and more likely), it is again an inhomogeneity in the right area itself, which causes a pseudo-Kondo-artifact. Apparently, the fabrication process induces inhomogeneities which are absent in  undamaged areas. We propose the following scenario to rationalize this finding: it is known the the PMMA resist can hardly be completely removed from the graphene area. While it seems that it does not induce significant inhomogeneities in undamaged graphene, it may well become visible after irradiation: After ion bombardment, the created defects (dangling bonds) react with ambient atmosphere (typically forming carboxy groups, among other possibilities). When, however, the area is covered with resist residues, this saturation of open bonds will happen very differently. This mechanism may induce significant inhomogeneities in ion treated areas, which may explain the pseudo-Kondo artifact in Fig. S1(b) and results in the light shaded area in Fig 1S(d). The similarity to the findings in \cite{fuhrer-kondo} is obvious.
\bibliography{Kondo_simulations}

\renewcommand{\thefigure}{S1}

\begin{figure*}[b!]
  \centering
	\includegraphics[width=178mm]{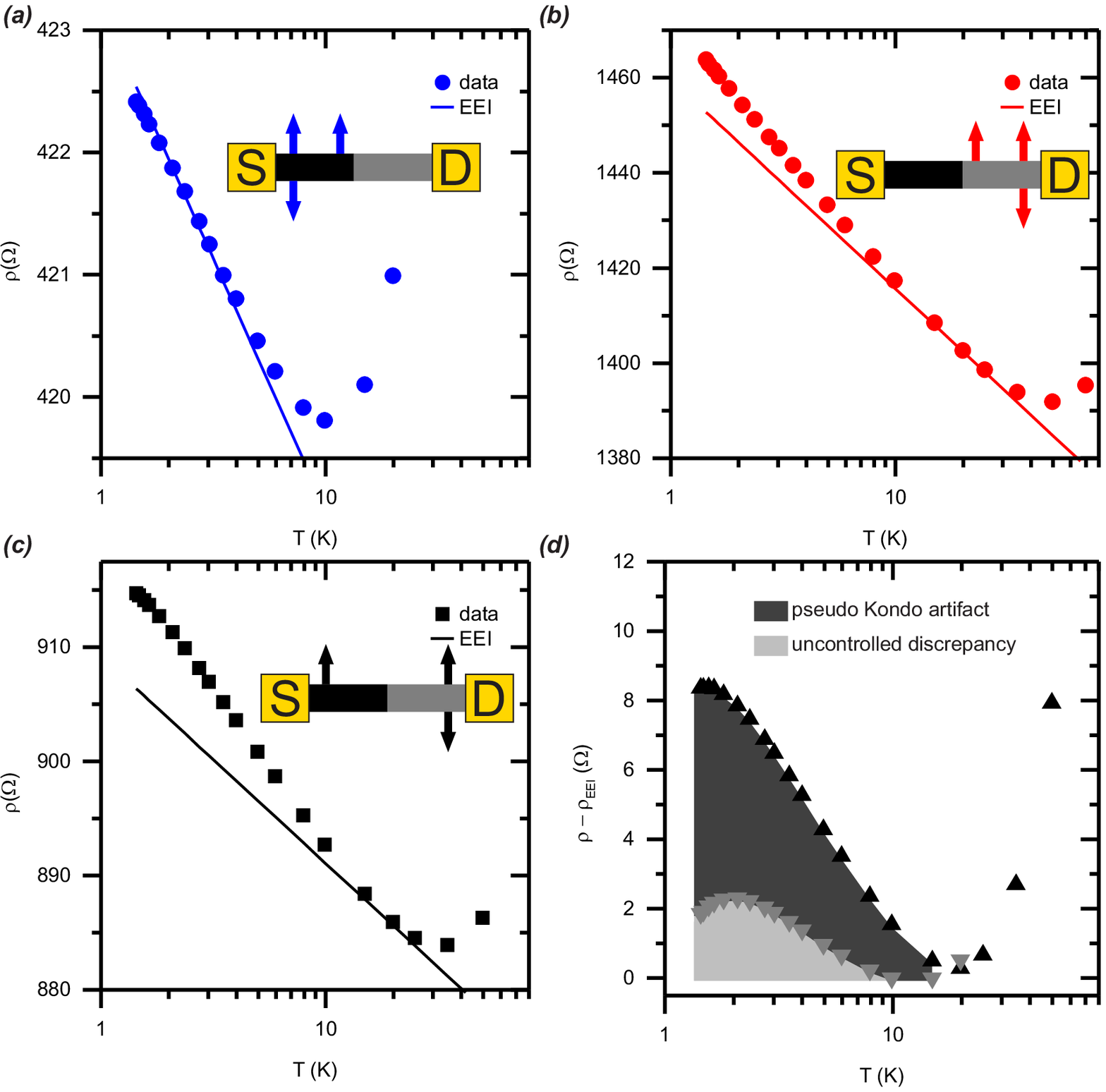}
	\caption{Artificially created inhomogeneous mono layer graphene (MLG) sample (see sketch in (a)--(c)) produced by ion beam bombardment of the right (grey) half of the structure. Hall--measurements were carried out independently in the as-grown left part and the ion-irradiated right part of the sample, as indicated graphically in figure (a)--(c). A magnetic field of $1\,$T was applied to suppress weak localization to a non-significant level. (a) Resistivity and EEI contribution of the as-grown MLG area (left part). (b) Resistivity and EEI contribution of the ion-beam-irradiated area (right part) assuming homogeneity. (c) Resistivity and EEI contribution of the complete structure. Hall--data is taken from the right part and homogeneity was assumed for the calculation of the EEI contribution. (d) Resulting difference if EEI is subtracted in each measurement assuming homogeneity. The dark shaded area denotes the pseudo Kondo artifact induced by EEI and inhomogenity. The light shaded area arises from the discrepancy in (b).}
\end{figure*}       
%

%
%